\documentclass[11pt]{article}

\usepackage[dvips]{graphicx}
\usepackage{amsfonts}
\usepackage{amssymb}

\newtheorem{theorem}{Theorem}
\newtheorem{proposition}{Proposition}
\newtheorem{remark}{Remark}

\newcommand{\R}{{\mathord{\mathbb R}}}
\newcommand{\Z}{{\mathord{\mathbb Z}}}
\newcommand{\N}{{\mathord{\mathbb N}}}
\newcommand{\KK}{{\mathord{\mathbb K}}}
\newcommand{\HH}{{\mathord{\mathbb H}}}
\newcommand{\MM}{{\mathord{\mathbb M}}}
\newcommand{\EE}{{\mathord{\mathbb E}}}

\newcommand{\MMS}{\MM^*}
\newcommand{\PMMS}{\partial\MM^*}
\newcommand{\beq}{\begin{equation}}
\newcommand{\eeq}{\end{equation}}
\newcommand{\bx}{{\bf x}}
\newcommand{\sds}{\strut\displaystyle}
\newcommand{\Dt}{\mbox{det\,}}
\newcommand{\Sg}{\mbox{sgn\,}}
\newcommand{\Cs}{\mbox{const.\,}}
\newcommand{\In}{\mbox{\,Inerdex\,}}
\newcommand{\Id}{\mbox{Ind\,}}
\newcommand{\Ai}{\mbox{Ai\,}}
\newcommand{\PP}{{\bf p}}
\newcommand{\UU}{{\bf 1}}
\def\staccrel#1#2{\mathrel{\mathop{#1}\limits_{#2}}}
\newcommand{\MC}{\stackrel{\scriptscriptstyle\hspace{-.9ex}\circ}{\MMS}}
\newcommand{\AC}{\staccrel{{\rm \scriptstyle after~the}}{{\rm \scriptstyle crossing}}}

\begin{document}

\title{\bf{Riemannian Geometrical Optics: Surface Waves in Diffractive Scattering}}

\author{\vspace{5pt} Enrico De Micheli$^{\dag}$, Giacomo Monti Bragadin$^{\ddag}$ and Giovanni Alberto Viano$^{*}$ \\
$^{\dag}$\small{Istituto di Cibernetica e Biofisica - Consiglio Nazionale delle Ricerche}\\[-5pt]
\small{Via De Marini, 6 - 16149 Genova, Italy. \vspace{8pt}} \\
$^{\ddag}$ \small{Dipartimento di Matematica - Universit\`a di Genova} \\[-5pt]
\small{Via Dodecaneso, 35 - 16146 Genova, Italy.\vspace{8pt}} \\
$^{*}$\small{Dipartimento di Fisica - Universit\`a di Genova} \\[-5pt]
\small{Istituto Nazionale di Fisica Nucleare - sez. di Genova} \\[-5pt]
\small{Via Dodecaneso, 33 - 16146 Genova, Italy.}
}

\date{}

\maketitle

\begin{abstract}
The geometrical diffraction theory, in the sense of Keller,
is here reconsidered as an obstacle problem in the Riemannian geometry. The first
result is the proof of the existence and the analysis of the main properties
of the {\it diffracted rays}, which follow from the non-uniqueness of the
Cauchy problem for geodesics in a Riemannian manifold with boundary. Then,
the axial caustic is here regarded as a conjugate locus, in the sense of the
Riemannian geometry, and the results of the Morse theory can be applied.
The methods of the algebraic topology allow us to introduce the homotopy classes
of diffracted rays. These geometrical results are related to the asymptotic
approximations of a solution of a boundary value problem for the
reduced wave equation. In particular, we connect the results of the Morse
theory to the Maslov construction, which is used to obtain the
uniformization of the asymptotic approximations. 
Then, the border of the diffracting body is the envelope of the diffracted
rays and, instead of the standard saddle point method, use is made of the
procedure of Chester, Friedman and Ursell to derive the
damping factors associated with the rays which propagate along the boundary.
Finally, the amplitude of the diffracted rays when the
diffracting body is an opaque sphere is explicitly calculated.
\end{abstract}

\newpage

\section{Introduction}
\label{introduction_section}
\begin{figure}[ht]
\vspace{-2.1in}
\centering
\includegraphics[width=7in]{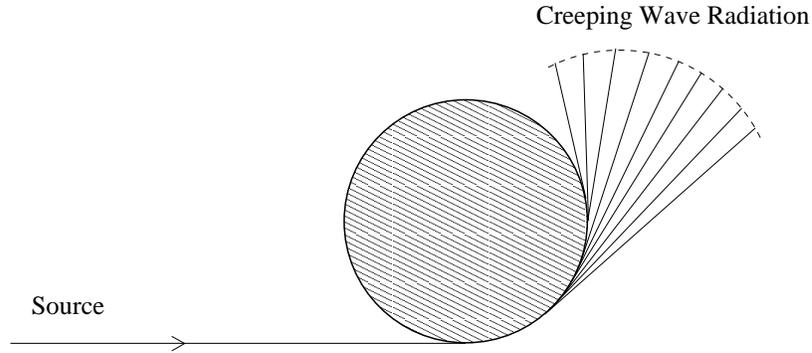}
\caption{
Geometric representation of the propagation of the creeping waves into the shadow of a
spherical obstacle.
}
\label{figura_1}
\end{figure}

Classical geometrical optics fails to explain the phenomenon of diffraction:
the existence of non-zero fields in the geometrical shadow. In several
papers \cite{Keller1,Keller2,Levy} Keller proposed an extension of classical geometrical 
optics to include diffraction (see also the book by Bouche, Molinet and Mittra 
\cite{Bouche} and references therein, where all these results have been collected and 
clearly exposed).
This modification basically consists in introducing new rays, called
{\it diffracted rays}, which account for the appearance of the light in the shadow. The
most clear and classical example of diffracted rays production is when a ray
grazes a boundary surface: the ray splits in two, one part keeps going as an ordinary ray, 
whereas the other part travels along the surface. At every point along its path this ray 
splits in two again: one part proceeds along the surface, and the other one leaves the surface along
the tangent to the surface itself (see Fig. \ref{figura_1}). 
Keller gives also a heuristic proof of the
existence of these diffracted rays which is based on an extension of the Fermat's principle
\cite{Keller1}. In spite of these efforts the concept of diffracted rays still remains
partially based on physical intuition. {\it The first aim of this paper is to put on
firm geometrical grounds the existence and the properties of the diffracted rays
when the diffracting body is a smooth, convex and opaque object.} To this
purpose the diffraction problem is here reconsidered as a Riemannian obstacle problem,
and then the diffracted rays arise as a consequence of the non-uniqueness of
the Cauchy problem for the geodesics at the boundary of the obstacle \cite{Alexander2,Alexander1} (i.e. the
diffracting body). Next, we are faced with the problem related to the caustic \cite{Berry1}, which is
composed of the obstacle boundary and of its axis (axial caustic). In
particular, since the latter can be regarded as a conjugate locus in the sense of the
differential geometry, all the classical results of the Morse theory 
\cite{Milnor1} must be formulated by taking into account the main geometrical 
peculiarity of the problem: the manifold we are considering has boundary.
Finally, the rays bending around the obstacle can be separated in
various homotopy classes by the use of the classical tools of algebraic
topology. All these geometrical questions will be analyzed in Section \ref{riemannian}.

Classical geometrical optics corresponds to the leading term of an asymptotic
expansion of a solution of a boundary value problem for the reduced wave
equation. This term, which is generally derived by the use of the stationary
phase method, gives approximations which have only a local
character: they are not uniform. In particular, these approximations fail on the
caustic; then a problem arises: how to patch up local solutions across the axial
caustic. It is well known in optics that after a ray crosses a caustic there is a
phase shift of $-\pi/2$.
It will be shown that this result can be derived in a very natural way by the use of the Maslov
construction \cite{Maslov1,Mishchenko1}, which effectively allows for linking the patchwork
occurring when the ray crosses the axial caustic and the geometrical analysis
of Section \ref{riemannian}. 

The surface of the diffracting body is the envelope of the diffracted rays:
it is a caustic and, consequently, the classical method of the stationary phase fails on it. 
By using a modified method of the stationary phase, due to 
Chester, Friedman and Ursell \cite{Chester}, we derive a countably infinite set of factors which
describe the damping of the creeping waves along the surface: these damping
factors depend on the obstacle curvature. These results, which lead to 
the Ludwig--Kravtsov \cite{Kravtsov1,Ludwig} uniform expansion at the caustic,
are briefly discussed in subsection \ref{airy}.

\noindent
From these introductory considerations the following geometrical ingredients emerge:
\begin{enumerate}
\item[i)] 
the non-uniqueness of the Cauchy problem at the boundary of the manifold in a
Riemannian obstacle problem;
\item[ii)]
the correspondence between the homotopy classes of the fundamental
group of the circle and the number of crossings through the axial caustic;
\item[iii)]
the Maslov phase-shift associated with the {\it crossing-number} of the ray;
\item[iv)]
the relationship between the curvature of the obstacle and the damping of the
creeping waves along the surface of the obstacle.
\end{enumerate}

In Section \ref{surface} a theory of the surface waves generated by diffraction,
that makes use of these geometrical properties, is presented.
Throughout this work we keep, as a typical example, the diffraction of the light by a
convex and opaque object, and, for the sake of simplicity, the light is represented
as a scalar. By using the same method, analogous results can be
obtained in sound diffraction (see the spectacular examples of creeping waves
in acoustic diffraction in Ref. \cite{Neubauer}) and, presumably, also in the diffraction 
of nuclear particles \cite{Fioravanti}.

A considerable part of the results that we obtain can be proved in a quite
general geometrical setting: this is the case, for instance, of the proof of the existence
of the diffracted rays, which derives solely from the non-uniqueness of the
Cauchy problem. But, for other results, we have to
restrict the class of obstacles to the surfaces of Besse type: i.e. the manifolds
all of whose geodesics are closed \cite{Besse}.
In particular, the sphere is a Besse surface. In this case all the results obtained by geometrical methods
can be easily compared with those obtained by using standard methods based on the expansion of the
amplitude in series (see subsection \ref{scattering}).
When the obstacle radius is large, compared to the wavelength, the series converge very
slowly, and the standard procedure suggests the use of the Watson resummation \cite{Sommerfeld,Watson}.
However, these formal manipulations do not shed light on the actual physical process.
Then, the geometrical approach is expected to be very useful in the investigation of more refined features, 
like {\it ripples} \cite{Nussenzveig2}, which are very sensitive to initial conditions and size parameters.

Finally, notice that our analysis will be limited to the geometrical theory of
surface waves, whose effects are dominant in a small backward angular region, as it will
be explained in subsection \ref{scattering}. The reader interested to a detailed analysis
of which effect is dominant in the various angular domains, and, accordingly,
to a systematic discussion of the transition regions, is referred to
Ref. \cite{Nussenzveig2} (in particular, Chapter 7 and Fig. 7.7).

\section{Riemannian Geometrical Optics}
\label{riemannian}
\subsection{Non-Uniqueness of Cauchy Problem in the Riemannian Manifold
with Boundary: the Diffracted Rays}
\label{non-uniqueness}
In the variational derivation of geometrical optics, in particular for the
laws of reflection and refraction, use is made of the Fermat's principle: the
paths of the reflected or refracted rays are stationary in the class of all
the paths that touch the boundary between two media at one point, assumed to be
an interior boundary point. To introduce the paths of the diffracted rays it is 
required a generalization of the Fermat's principle, extended to include points, 
as well as arcs, lying on the boundary \cite{Nussenzveig2}. In our analysis, 
instead of the Fermat's, we use the Jacobi form of the principle of least action \cite{Goldstein}
which concerns with the path of the system point rather than with its motion in time. 
More precisely, the Jacobi principle states: if there are no forces acting on the body,
then the system point travels along the shortest path length in the configuration
space. Moving from mechanics to optics, Riemannian geometrical optics can be rested on
the Jacobi principle, formulated as follows: {\it the light rays travel along geodesics.}

In this context the diffraction by a convex, smooth and opaque object can be
reconsidered as a Riemannian obstacle problem: the object is regarded as an
obstacle which a geodesic can bend around, or which a geodesic can end at.
Let $\KK$ denote the obstacle which is embedded in a complete $n$-dimensional
Riemannian manifold $(\HH, g)$, where $g$ is the metric of $\HH$ and $n \geq 3$.
Let us introduce the space $\MM = \HH \backslash \overline{\KK}$, 
where the obstacle $\KK$ is an open connected subset of $\HH$, with regular
boundary $\partial \KK$ and compact closure $\overline{\KK} = \KK \cup \partial \KK$.
Although most of the results illustrated below
hold true in a very general setting, in the following we keep very often, as
a typical example, $\HH \equiv \R^3$ endowed with the euclidean metric.
Finally, we are led to consider the space $\MMS = \MM \cup \partial \KK
\left( = \HH \backslash \KK \right)$, that has the structure of a
manifold with boundary. 

Now, two kinds of difficulties arise: the first one concerns
geodesic completeness, that is the possibility to extend every geodesic
infinitely and in a unique way; this uniqueness is indeed missing in $\MMS$ at the points of the boundary. 
The second difficulty is related to the necessity of finding suitable coordinates at the boundary
which allow for using the ordinary tools of the differential geometry, 
e.g. for writing the equations of the geodesics.

Concerning the first point, the lack of geodesic completeness can be
treated by introducing the notion of {\it geodesic terminal} (see Ref. \cite{Plaut}) to
represent a point where a geodesic stops. Following Plaut \cite{Plaut}, it can be proved
that $\MMS$ is the completion of $\MM$ by observing that the set
${\cal I}$ of the geodesic terminals is nowhere dense and 
$\PMMS = \partial \KK = {\cal I}$.

Regarding the second point, it is convenient to model 
the manifold with boundary on the half--space 
$\R^n_+ = \left\{ (x_1,...,x_n) \in \R^n | x_n \geq 0 \right\} \subset \R^n$,
where $\R^{n-1}_0$ denotes the boundary $x_n = 0$ of $\R^n_+$.
Thus, for a manifold with boundary, there exists an atlas $\{{\cal U}^\alpha\}$, 
with local coordinates $(x_1^\alpha,...,x_n^\alpha)$,
such that in any chart we have the strict inequality 
$x^\alpha_n > 0$ at the interior points, and $x^\alpha_n = 0$
at the boundary points. The set $\PMMS$ of the boundary points is a smooth
manifold of dimension $(n-1)$. 

In a Riemannian manifold without boundary, the geodesics
are the solutions of a system of differential equations with suitable Cauchy
conditions. The main result is the theorem which guarantees existence,
uniqueness and smoothness of the solution for the Cauchy problem of this
system. The case of Riemannian manifold with boundary is different from the
classical one for what concerns uniqueness and smoothness of the solutions. 

In order to write the equation of a geodesic
$\gamma$ of $\MMS$, we introduce suitable coordinates $x_i$ $(i=1,\ldots,n)$
adapted to the boundary, called {\it geodesic boundary coordinates} \cite{Alexander2,Alexander1},
with $x_n$ defined as the distance from the boundary $\PMMS$;
then starting with arbitrary coordinates $x_i,~i<n$ on $\PMMS$,
we extend them to be constant on ordinary geodesics normal to $\PMMS$. 
Let $\Gamma_{ijk}$ denote the Christoffel symbols of the Levi Civita connection of
$\MMS$, and $\chi$ be the normal curvature of $\PMMS$ 
in the direction of $\gamma$. Then, we have:
\beq
\label{uno}
\chi=-\sum_{i,j<n} \dot{x}_i \, \dot{x}_j \, \Gamma_{ijn}.
\eeq
When $\gamma$ is not in $\PMMS$ the same expression occurs in the differential 
equations for $\gamma$, so that if we define $\chi=0$ off the boundary 
segments, then the differential equations can be modified to cover, in an integral 
sense, all points of $\gamma$, as follows \cite{Alexander2,Alexander1}:
\begin{eqnarray}
\sds \ddot{x}_k & = & -\sum_{i,j} \dot{x}_i \, \dot{x}_j \, \Gamma_{ijk},~~~~~(k<n), \label{duea}\\
\sds \ddot{x}_n & = & -\chi-\sum_{i,j<n} \dot{x}_i \, \dot{x}_j \, \Gamma_{ijn}. \label{dueb}
\end{eqnarray}

Equations (\ref{duea}, \ref{dueb}) can be implemented with the Cauchy conditions by providing
the values of $\bx(t)$ and $\dot{\bx}(t)\,\, (\bx\equiv(x_1,..., x_n))$ when $t=0$:
\begin{eqnarray}
x_i(0)       & = & b_i, ~~~~~ (i<n), ~~~ x_n(0)=0, \label{trea} \\
\dot{x}_i(0) & = & v_i, ~~~~~ (i<n), ~~~ \dot{x}_n(0)=0, \label{treb}
\end{eqnarray}
where $(b_1,...,b_n)$ are the coordinates of the point $b$ belonging to
the boundary, and $(v_1,...,v_n)$ are the components of the tangent vector at
$b$.

Now, let us consider the following cases:
\begin{enumerate}
\item [a)] $x_n \equiv 0$. In view of equalities (\ref{uno}) and (\ref{dueb}), 
we remain with system (\ref{duea}) composed by $(n-1)$ differential
equations that describe a geodesic on the $(n-1)$ dimensional smooth manifold
$\PMMS$ (i.e. $x_n=0$), which is a manifold without boundary.
In this case the standard theorem on the existence, uniqueness and smoothness
of the geodesics satisfying the given Cauchy conditions holds true. We have a
$C^{\infty}$-class geodesic which is completely contained on the boundary.
\item[b)] $x_n (t)=0$ if $t\leq 0$, and $x_n(t)>0$ if $t>0$.
In view of the fact that $\chi=0$ off the boundary segments,
eqs. (\ref{duea}, \ref{dueb}) describe, for $t>0$, a geodesic segment belonging to the
interior of $\MMS$ (i.e. $\MC$). Moreover, eqs. (\ref{duea}, \ref{dueb})
describe, for $t<0$, a geodesic segment belonging to the boundary. Notice
that the r.h.s. of (\ref{duea}) is continuous, so that (\ref{duea}) holds everywhere as it
is. If $\chi \neq 0$ on boundary segments, from Eq. (\ref{dueb}) it can be easily
seen that $\ddot{x}_n$ fails to be continuous at the point $b$ $(t=0)$.
Nevertheless, in view of the Cauchy conditions (i.e.
$x_n(0)=\dot{x}_n(0)=0$), the two geodesic segments (one belonging to the
boundary, the other belonging to the interior of $\MMS$) glue
necessarily at $b$, and give rise to a geodesic of
$\MMS$ of class $C^1$ at the transitions point $b$.
\item[c)]
$x_n(t)=0$, if $t\geq 0$, $x_n(t)>0$ if $t<0$.
The analysis is analogous to the previous one at point (b). We can
say that two geodesic segments (one belonging to the interior of $\MMS$,
the other one to the boundary) glue together at $b$ and give
rise to a geodesic of $\MMS$ of class $C^1$ at the transition point $b$.
\item[d)]
$x_n(t)=0$ if and only if $t=0$. Following
considerations strictly analogous to those developed above, we can say that
eqs. (\ref{duea}, \ref{dueb}) describes two geodesic segments which belong to the interior of
$\MMS$, and have only one point of contact with the boundary at $b$. 
Therefore, gluing at $b$ the two geodesic segments, we obtain a geodesic of $\MMS$ of class
$C^{\infty}$.
\end{enumerate}

Since we supposed the obstacle to be compact with smooth and convex
boundary, the analysis performed above is exhaustive.

Summarizing, if $\chi \neq 0$ on the boundary segments, 
we can glue together, at the point $b\in \PMMS$, a geodesic segment belonging to the boundary with
a geodesic segment belonging to the interior of $\MMS$
(cases (b) and (c)), and, in these case, we have a
geodesic of $\MMS$, which is of class $C^1$ at the transition point
$b$. Thus, we have obtained the diffracted rays, which are precisely the
geodesics of $\MMS$ of this type. The situation considered in case
(d) is of minor interest for our analysis, since those geodesics
correspond to rays which are not diffracted even if they touch the
obstacle.

As a consequence of the non-uniqueness of the Cauchy problem at the boundary of
the manifold, that can be made transparent by the use of equations (\ref{duea}) and (\ref{dueb}),
we can conclude by saying that at each point of the boundary we have a
bifurcation: the ray splits in two, one part continues as an ordinary ray
(diffracted ray of class $C^1$ at the splitting point), the other part
travels along the surface as a geodesic of the boundary.

Finally, let us observe that the contact point $b$ on the boundary is
necessarily an elliptic point in view of the assumptions of convexity and
compactness made about the obstacle.

\subsection{Conjugate Points (Axial Caustic) and Morse theorem in a Riemannian Manifold with Boundary}
\label{conjugate}
First, let us consider a complete Riemannian manifold $\MM$ without
boundary. Given a geodesic $\gamma=\gamma(t)$, $0\leq t\leq 1$, consider a
geodesic variation of $\gamma$, that is a one-parameter family of geodesics
$\gamma_s=\gamma_s (t)$, $(-\epsilon < s < \epsilon)$ such that $\gamma_0=\gamma$. 
For each fixed $s$, $\gamma_s (t)$ describes a geodesic when $t$ varies from $0$
to $1$. Each variation gives rise to an infinitesimal variation, that is, a
certain vector field defined along $\gamma$. The points $\gamma(0)$ and
$\gamma(1)$ of $\gamma$ are said to be conjugate if there is a variation
$\gamma_s$ which induces an infinitesimal variation vanishing at $t=0$ and
$t=1$ (see Ref. \cite{Kobayashi}). 

Set $f(s,t)=\gamma_s(t)$, and denote by $D/{\partial t}$ and $D/{\partial s}$ the
covariant differentiation with respect to $t$ and $s$, respectively. 
Denote by $R$ the curvature transformation determined by
$\partial f/\partial t$ and $\partial f/\partial s$, i.e.
\beq
\label{cinque}
R \left (\frac{\partial f}{\partial t},\frac{\partial f}{\partial s}\right)\frac{\partial f}{\partial t}=
\left (\frac{D}{\partial s} \, \frac{D}{\partial t}-\frac{D}{\partial t} \,
\frac{D}{\partial s}\right ) \frac{\partial f}{\partial t}.
\eeq
\noindent
A vector field $Y$ is a Jacobi field along $\gamma$ if satisfies the following differential equation:
\beq
\label{nove}
J\, Y \equiv Y^{''} + R \, (\gamma^{'}, Y) \, \gamma^{'}=0, 
\eeq
\noindent
where $\gamma^{'} \equiv d\gamma/dt$, $Y^{''}$ is the second covariant
derivative of $Y$ along $\gamma$, and $R\,(\gamma^{'}, Y)$ is the curvature
transformation defined by equality (\ref{cinque}). Two points $p, q \in \MM$ are
conjugate along $\gamma$ if there exists a non-trivial Jacobi field $Y$ along
$\gamma$ such that $Y(p)=Y(q)=0$. 
Finally, the multiplicity of the pair of conjugate points $p,q\in \gamma$ is given by the
dimension of the vector space of the linearly independent Jacobi fields, along
$\gamma$, that vanish at $p$ and $q$.

Let us now introduce the infinite dimensional space $\Omega (\MM; p,q)$ of 
piecewise differentiable paths $c$ connecting on $\MM$ the point
$p$ with the point $q$: i.e. let $p$ and $q$ be two fixed points on $\MM$,
and $c:[0,1]\rightarrow \MM$ be a piecewise differentiable path such
that $c\,(0)=p$, $c\,(1)=q$. To any element $c\in\Omega (\MM; p,q)$ 
we associate an infinite dimensional vector space
$T_c\Omega$, which can be identified with the space tangent to $\Omega$ in a
{\it point} $c$. More precisely $T_c\Omega$ is the vector space composed by all
those fields of piecewise differentiable vectors $V$ along the path $c$, such
that $V(0)=0$, $V(1)=0$. Next, we introduce the {\it energy-functional}
\beq
\label{dieci}
E(c)=\int_0^1 \: \left | \frac{dc}{dt}\right |^2 \, dt =
\int_0^1 \, g_{ij}\, \dot{x}^i\, \dot{x}^j \, dt,
\eeq
\noindent
where $g_{ij}$ is the metric of $\MM$, and $(x^1,....,x^n)$ are local
coordinates on $\MM$. From the first variation of functional $E$ we can
deduce that the extremals of functional $E(c)$ are represented by the
geodesics $c\,(t)=\gamma(t)$, parametrized by $t$.
Let us now consider the second variation $E_{**}$ of the functional
$E$ along the geodesic $\gamma$ (it will be called
the {\it hessian} of $E$), and let $\lambda$ be the index of the hessian, that is
the largest dimension of the subspace of $T_{\gamma}\Omega$
on which $E_{**}$ is definite negative. One can then state the following Theorem:

\begin{theorem}{{\rm(Morse Index Theorem \cite{Milnor1})}}
The index $\lambda$ of the hessian $E_{**}$ equals the number of points belonging
to $\gamma(t)$ which are conjugate to the initial point $p = \gamma(0)$, each one
counted with its multiplicity.
\end{theorem}

Now let us go back to our main concern, the Riemannian manifold with boundary $\MMS$.
Following Alexander \cite{Alexander3}, we define as a Jacobi field any vector field $Y$ along
the geodesic $\gamma$ (whose boundary contact intervals have positive measure), which
satisfies the following conditions: $Y$ is continuous, it is a Jacobi field of both
$\MC$ and $\PMMS$ along every segment of $\gamma$, 
and at each endpoint $t_i$ of a contact interval it satisfies the following equation:
\beq
\label{undici}
P\left (\frac{D}{\partial t} \, Y \right ) (t_i^-)
= P \left ( \frac{D}{\partial t} \, Y \right ) (t_i^+),
\eeq
\noindent
where $P$ denotes orthogonal projections onto the hyperplane tangent to
$\PMMS$. In the Riemannian manifold with boundary we are forced
to consider geodesics which lie both on and off $\PMMS$, and that are
merely $C^1$ at the transition points (see the previous subsection).
However, it is still possible to define a hessian form which is given simply by the
sum of the classical formulae for $\PMMS$ on contact intervals, and
for the interior of $\MMS$ (i.e. $\MC$) on interior intervals.
Furthermore, we call $\gamma(0)$ and $\gamma(1)$ conjugate if there is a non trivial
Jacobi field $Y$ along $\gamma$ whose limits at the endpoints vanish. Finally,
the Morse index theorem can be extended to the Riemannian manifold with 
boundary. To this purpose, it is convenient to introduce the so-called
{\it regular geodesics} in the sense of Alexander. Following Ref. \cite{Alexander3},
a geodesic $\gamma$ is regular if:
\begin{enumerate}
\item[a)] all boundary-contact intervals of $\gamma(t)$ have positive measure;
\item[b)] the points of arrival of $\gamma(t)$ at $\PMMS$ are not
conjugate to the initial point $\gamma(0)$.
\end{enumerate}
Then we can formulate the extended Morse index theorem as follows:
\begin{theorem}[{\rm Morse Index Theorem for Riemannian Manifolds with Boundary \cite{Alexander3}}] 
Let $\gamma$ be a regular geodesic; then the index $\lambda$ of 
$E_{**}$ is finite and equal to the number of points belonging
to $\gamma(t)$ conjugate to the initial point $\gamma(0)$ ($0 \leq t \leq 1$), 
counted with their multiplicities. 
\end{theorem}

It still remains to check if the diffracted rays are regular geodesics. First
of all we suppose that the source of light is located in a point 
$p_0 \in \MC$ (point source); then the diffracted rays
which we are considering connect the point $p_0$
and a point $q \in \MC$, placed at the exterior to the obstacle.
If we assume that $\HH = \R^3$ (equipped with euclidean metric), then the diffracted rays consist
of a straight line segment from $p_0$ to the body surface, of a geodesic along
the surface, and a straight line segment from the body to $q$. Now, in view of
the fact that these rays undergo diffraction, they cannot be simply tangent to the obstacle, but
the contact-interval must be a geodesic segment of positive measure. Therefore
condition (a) is satisfied. Concerning condition (b), we simply note
that it is certainly satisfied in view of the fact that we consider obstacles
formed by a convex and compact body embedded in $\R^3$, and the latter
does not have conjugate points.

\subsection{Morse Index and Homotopy Classes of Diffracted Rays}
\label{morse}
Hereafter the analysis will be limited to obstacles
represented by manifolds all of whose geodesics are closed (i.e. Besse manifolds \cite{Besse}).
Merely for the sake of simplicity, henceforth we shall consider as obstacles spherical balls
embedded in $\R^3$: i.e $\partial \KK \equiv S^2$, equipped with standard metric. 
However, in view of the homotopic invariance, the main result of this section 
(see next (\ref{quattordicia}, \ref{quattordicib}))
holds true for any convex and compact Besse type manifold. \\
Consider the axis of symmetry $A$ of the obstacle $\KK$ passing through the point $p_0 \in \MC$,
where it is located the source of light, and denote by $A_-$ the portion of this
axis lying in the illuminated region (i.e. the same side of the light source), and
by $A_+$ the portion of the axis lying in the shadow. The axis $A$ equals
$A_- \cup D \cup A_+$, where $D$ is the diameter of the sphere.
Let us note that all the points of $\MMS$
that do not belong to $A_+$ are connected to $p_0$ by only one geodesic of minimal length, whereas the points
$q \in A_+$ are connected to $p_0$ by a continuum of geodesics of minimal length that
can be obtained as the intersections of $\MMS$ with planes
passing through the axis $A$. By rotating these planes, and keeping $p_0$ and $q \in A_+$
fixed, we obtain a variation vector field which is a Jacobi field vanishing at $p_0$ and $q$.
Thus, we can conclude that $p_0$ and $q \in A_+$ are conjugate
with multiplicity one because the possible rotations are only
along one direction. Then, in view of the extended form of the Morse
Index Theorem, we can state that the index $\lambda$ of $E_{**}$
jumps by one when the geodesic $\gamma(t)$, whose initial point is $\gamma(0)=p_0$, crosses $A_+$.

Now let us focus our attention on the points $q \in \MMS \backslash (A_+ \cup A_-)$
(i.e. which do not lie on the axis of the obstacle, see Fig. \ref{figura_2}). Consider the space
$X_{p_0 q} = (\EE^2)_q \cap \MMS$, where $(\EE^2)_q$ is the plane uniquely
determined by the axis of the obstacle through $p_0$ and the point $q$. 
$X_{p_0 q}$ (which will be denoted hereafter simply by $X$) is an
arcwise connected space, whose boundary in $(\EE^2)_q$ is a circle $S$ which is a
deformation retract of $X$. In view of these facts the fundamental group
$\pi_1(X;p_0)$ does not depend on the base point $p_0$ and is isomorphic
to $\pi_1(S^1; \UU)$, where $\UU$, described in a convenient system by the
coordinates $(0,1)$ (see Fig. \ref{figura_2}), represents the contact point of
$A_+$ with $S^2$: $\pi_1(X;p_0) \simeq \pi_1(S^1;\UU) \simeq \Z$.
Let us consider, within the fundamental groupoid $\Pi_1(X)$ of $X$, the set
$\Pi_1(X;p_0,q)$ of paths in $X$ connecting $p_0$ with $q$, modulo
homotopy with fixed end-points. We can then formulate the following statements:

\begin{figure}[htb]
\vspace{-5.2in}
\centering
\includegraphics[width=9in]{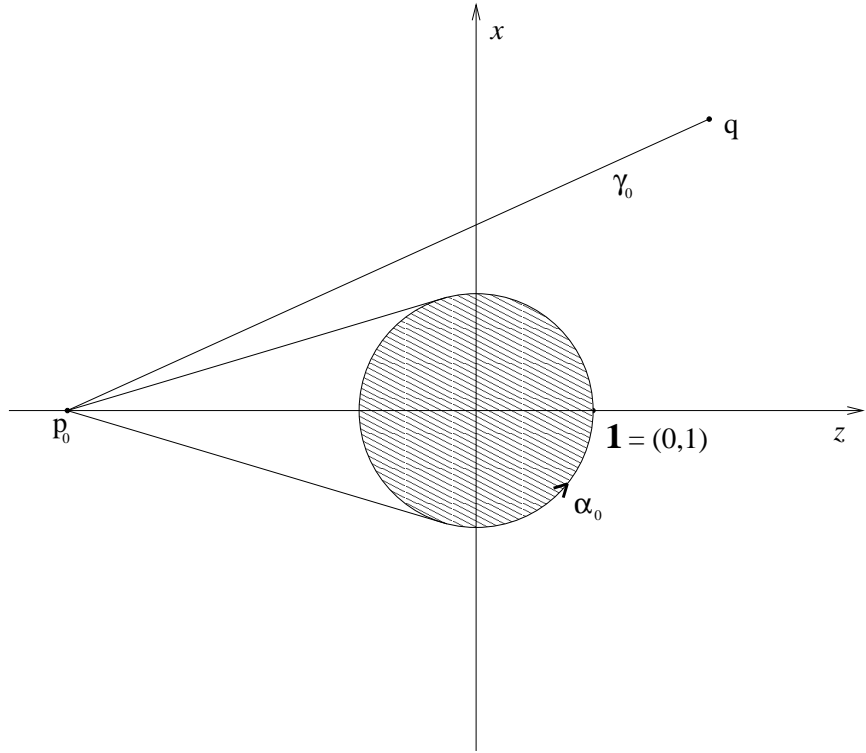}
\caption{
Geometry of the problem in the case of spherical obstacle. $\gamma_0$ is the unique
geodesic of minimal length when the point $q$ does not belong to the axial caustic (the $z$-axis).
$\alpha_0$ is a counterclockwise loop at $p_0$.
}
\label{figura_2}
\end{figure}

\begin{proposition}
\label{pro:1}
Each element of $\pi_1(X;p_0)$ is a homotopy class $[\alpha]$,
with fixed endpoints, of a certain loop $\alpha:[0,1]\rightarrow X$, in the space
$X$, starting and ending at the point $p_0$.
\end{proposition}

\begin{proposition} 
\label{pro:2}
Each path $c_0$ from $p_0$ to $q \in \MMS \backslash
(A_+ \cup A_-)$, determines a one-to-one correspondence $W$ between
$\pi_1(X;p_0)$ and the set $\Pi_1(X; p_0,q)$.
Such a correspondence can be constructed as:
$\forall\,[c] \in \Pi_1(X;p_0,q):[c]\longrightarrow [c \star c_0^{-1}]\in \pi_1(X;p_0)$,
where the symbol '$\star$' denotes the concatenation of paths, and $c_0^{-1}$
denotes the reverse path of $c_0$. 
\end{proposition}

\begin{proposition}
\label{pro:3}
i) In each homotopy  class of $\pi_1(X;p_0)$ there is precisely one element
of minimal length.
ii) In each homotopy class of $\Pi_1(X;p_0,q)$, $q\in \MMS \backslash 
(A_+ \cup A_-)$, there is precisely one geodesic.
\end{proposition}

Because of Propositions \ref{pro:2} and \ref{pro:3} we can establish a bijective correspondence $w$
between geodesics from $p_0$ to $q$ (fixed) and integral numbers. Let us make
precise our choices: 
\begin{enumerate}
\item[{\sl a})] $q$ is fixed in $X\backslash (A_+ \cup A_-)$; we introduce in $(\EE^2)_q$
a reference system so that $p_0$ belongs to the negative part of the $z$-axis,
and the value of the coordinate $x$ of the point $q$ (i.e. $x_q$) is positive
(see Fig. \ref{figura_2});
\item[{\sl b})] $\gamma_0$ is the geodesic from $p_0$ to $q$ of minimal length;
\item[{\sl c})] $\alpha_0$ is a loop in $X$ at $p_0$, such that:
\begin{enumerate}
\item[{\sl c'})] $[\alpha_0]$ is a generator of $\pi_1(X;p_0)$ (establishing
an isomorphism $\pi_1(X;p_0) \simeq \Z$);
\item[{\sl c"})] $\alpha_0$ turns in counterclockwise sense around the obstacle (see Fig. \ref{figura_2}).
\end{enumerate}
\end{enumerate}
Now, the correspondence $w$ maps the geodesic $\gamma$ into the number
$w(\gamma)$ such that $[\gamma \star \gamma_0^{-1}] = w(\gamma) \, [\alpha_0]$.
\noindent
Note that $w(\gamma)$ is the winding number of the loop $[\gamma\star\gamma_0^{-1}]$
determined by $\gamma$ (with respect to the chosen generator $[\alpha_0]$).

We can also characterize the geodesics (from $p_0$ to $q$) by a natural number
$n(\gamma)$, which we call the {\it crossing-number} of $\gamma$, having a clearer
geometric interpretation: $n(\gamma)$ counts the number of crossings of $\gamma$
across the $z$-axis. For instance, $n(\gamma_0) = 0$.
It is easy to see that the winding number determines the crossing-number
through the following bijective correspondence:
\begin{eqnarray}
\Z & ~~~\longrightarrow ~~~ & \hspace{0.5cm} \N \nonumber \\
m        & ~~~\longrightarrow ~~~ & (2m-1)~~~~~~~~(m>0) \label{quattordicia} \\
m        & ~~~\longrightarrow ~~~ & \hspace{0.25cm} -2m \,~~~~~~~~~~~(m \leq 0) \label{quattordicib}
\end{eqnarray}
so that our previous statement on the bijective correspondence between the geodesics and
their crossing-number is established.

\section{Surface Waves in Diffractive Scattering}
\label{surface}
\subsection{Creeping Waves on the Sphere}
\label{creeping}
In this subsection we consider a Riemannian manifold without boundary $\MM^n$;
$\Omega$ is a chart of $\MM^n$, $\bx =(x_1,...,x_n)$ are the local coordinates in $\Omega$ 
and $g_{ij}(\bx)$ is the metric tensor. As usual $g=|\Dt (g_{ij})|$ and 
$g^{ij}=g_{ij}^{-1}$.
Let us consider the Helmholtz equation
\beq
\label{quindici}
\Delta_2 u + k^2 u = 0,
\eeq
\noindent
where $\Delta_2$ is the Laplace-Beltrami operator which, for a function 
$u \in C^{\infty}(\MM^n)$, reads
\beq
\label{sedici}
\Delta_2 u = \frac{1}{\sqrt g} \, \sum_{i=1}^n \, \frac{\partial}{\partial x_i}
\left ( \sum_{j=1}^n g^{ij} \, \sqrt g \, \frac{\partial}{\partial x_j} \right ) u.
\eeq
\noindent
Now, we look for a solution of Eq. (\ref{quindici}) of the following form:
\beq
\label{diciassette}
u(\bx,k)=\int A(\bx,\beta)~e^{ik\Phi(\bx,\beta)}~d\beta.
\eeq
\noindent
The principal contribution to $u(\bx,k)$, as $k\rightarrow +\infty$, corresponds to the
stationary points of $\Phi$, in the neighbourhoods of which the exponential
$e^{ik\Phi}$ ceases to oscillate rapidly. These stationary points can be obtained
from the equation $\partial\Phi(\bx,\beta)/\partial\beta = 0$ 
(provided that $\partial^2\Phi/\partial\beta^2\neq 0$). 
Let us now suppose that for each point 
$\bx = (x_1,...,x_n)$ $\Phi$ has only one stationary point; then the following
asymptotic expansion of $u$, as $k\rightarrow\infty$, is valid \cite{Maslov1}:
\beq
\label{diciotto}
u(\bx,k)\, \simeq\, e^{ik\Phi(\bx,\beta_0)}\sum_{m=0}^\infty\frac{A_m}{(ik)^m},
\eeq
\noindent
\noindent
where $\beta_0$ is the unique stationary point of $\Phi$.
The leading term of expansion (\ref{diciotto}) is: 
\beq
\label{diciannove}
u(\bx,k)=A_0(\bx)\, e^{ik\Phi(\bx,\beta_0)},
\eeq
\noindent
where
\beq
\label{venti}
A_0(\bx)=A(\bx,\beta_0)\left (\left | \frac{\partial^2\Phi}{\partial\beta^2}\right |^{-\frac{1}{2}}
\right )_{\!\!\beta=\beta_0} \exp \left \{ {i\,\frac{\pi}{4} {\,\,\Sg} \left (\frac{\partial^2\Phi}
{\partial\beta^2}\right )_{\!\beta=\beta_0}} \right \}.
\eeq
\noindent
By substituting the leading term (which will be written hereafter as $A\exp\{ik\Phi\}$, omitting
the index zero) into equation (\ref{quindici}), collecting the powers of $(ik)$, and, finally, 
equating to zero their coefficients, two equations are obtained: the eikonal (or
Hamilton-Jacobi) equation
\beq
\label{ventuno}
g^{ij}\frac{\partial\Phi}{\partial x_i}\, \frac{\partial\Phi}{\partial x_j} = 1,
\eeq
\noindent
and the transport equation
\beq
\label{ventidue}
\frac{1}{\sqrt g}\, \sum_{i=1}^n\frac{\partial}{\partial x_i} \left \{ \sqrt g
\left ( A^2\sum_{j=1}^n g^{ij} \, \frac{\partial\Phi}{\partial x_j}\right )\right \} = 0,
\eeq
\noindent
whose physical meaning is the conservation of the current density.

We can switch from wave to ray description by writing the eikonal equation
in the form of a Hamilton-like system of differential equations by setting
\begin{eqnarray}
\sds p_i & = & \frac{\partial\Phi}{\partial x_i}, \label{ventitrea} \\
\sds F   & = & \frac{1}{2} \left ( g^{ij}\, p_i \,  p_j - 1 \right ), \label{ventitreb}
\end{eqnarray}
and obtaining the characteristic system
\begin{eqnarray}
\sds \frac{dx_i}{d\tau} & = & g^{ij} \, p_j, \label{ventiquattroa} \\
\sds \frac{dp_i}{d\tau} & = & -\frac{\partial F}{\partial x_i} = -\frac{1}{2} \: p_i \, p_j \,
\frac{\partial g^{ij}}{\partial x_i}, \label{ventiquattrob}
\end{eqnarray}
where $\tau$ is a running parameter along the ray emerging from the surface $\Phi=$ const.

Whenever the amplitude $A$ becomes infinite, approximation
(\ref{diciannove}) fails, and, consequently, such an approximation
holds true only locally. Thus, we are faced with the problem
of passing from a local to a global approximation, i.e. a solution in the whole space. 
The strategy for constructing a solution in the whole space consists in patching up local
solutions by means of the so called {\it Maslov-indexes} in a way that will be
illustrated below in the specific case of $\MM^n=S^2$.

Let the unit sphere be described by the angular coordinates $\theta$ and $\phi$. 
The matrix elements $g_{ij}$ have the following values: $g_{11}=1$, $g_{22}=\sin^2\theta$, 
$g_{12}=g_{21}=0$. In addition, we
assume that the phase $\Phi$ and the amplitude $A$ do not depend on the
angle $\phi$. Then, eqs. (\ref{ventuno}) and (\ref{ventidue})
become, respectively
\begin{eqnarray}
\label{venticinque}
&&\left (\frac{d\Phi}{d\theta}\right )^2=1, \\
\label{ventisei}
&&\frac{1}{|\sin\theta |}\left\{\frac{d}{d\theta}\left (|\sin\theta |\: A^2 \: \frac{d\Phi}{d\theta}\right )\right\}=0,
~~~~~ \theta \neq n\pi.
\end{eqnarray}
\noindent
From Eq. (\ref{venticinque}) we have: $\Phi=\pm\theta + \Cs$, and from Eq. (\ref{ventisei}) 
we obtain: $A\,(\theta)=\Cs \cdot |\sin\theta|^{-1/2},\, \theta \neq n\pi$.
Therefore, approximation (\ref{diciannove}) becomes
\beq
\label{ventotto}
u\,(\theta,k)=\frac{\Cs}{\sqrt{|\sin\theta |}} ~ e^{\pm ik\theta},~~~~~~~~~ \theta \neq n \pi,
\eeq
\noindent
where the terms $\exp\{\pm ik\theta\}$ represent waves traveling counterclockwise
($\exp\{ik\theta\}$) or clockwise ($\exp\{-ik\theta\}$) around the unit sphere.
Since the approximation (\ref{ventotto}) fails at
$\theta=n\pi$, ($n=0,1,2...$), we have to consider the problem of patching
these approximations when the surface rays
cross the antipodal points $\theta=0,\pi$, which are conjugate points in the
sense of the Morse theory (see Section \ref{morse}).
This difficulty will be overcome by the use of the Maslov construction \cite{Delos,Maslov2,Maslov1,Mishchenko1}.

In order to illustrate the Maslov scheme in the case $\MM^n \equiv S^2$,
we reconsider the problem of the waves creeping around the unit sphere in a more
general setting.
The Hamilton-Jacobi equation (\ref{ventuno}) is rewritten in the form:
\beq
\label{ventinove}
H=g^{ij}\, p_i \, p_j = p_\theta^2 + \frac{1}{\sin^2\theta} \: p_\phi^2 = 1.
\eeq
\noindent
In order to find the first integrals $h$ of the system, we equal to zero the 
Poisson brackets: $\{h,H\}=0$, to get:
\begin{eqnarray}
\sds p_\phi & = & c_1, \label{trentaa} \\
\sds p_\theta^2 & + & \frac{c_1^2}{\sin^2\theta}=1. \label{trentab}
\end{eqnarray}
\noindent
Furthermore, we have:
\beq
\label{trentuno}
\Phi_\pm (\theta,\phi) = \int (p_\phi \, d\phi + p_\theta\, d\theta ) = c_1 \phi \pm \Phi_1(\theta),
\eeq
\noindent
(notice that $p_\theta = \pm \{1-(c_1^2 / \sin^2\theta)\}^{1/2}$).
Equation (\ref{trentab}) defines a smooth curve in the domain $0<\theta<\pi$,
$p_\theta\in\R$, which is diffeomorphic to the circle. The points where the tangent
to the curve is vertical have coordinates $(\theta_0,0)$, $(\pi-\theta_0,0)$,
with $\theta_0=\sin^{-1}(c_1)$, $c_1>0$. The cycles of singularities are
given by the equations (see Ref. \cite{Maslov1}, Fig. 16):
\begin{eqnarray}
p_\phi & = & c_1,~~p_\theta=0;~~~~\theta=\theta_0,~~0\leq\phi\leq 2\pi, \label{trentaduea} \\
p_\phi & = & c_1,~~p_\theta=0;~~~~\theta=\pi - \theta_0,~~0\leq\phi\leq 2\pi. \label{trentadueb}
\end{eqnarray}
Then, the neighbourhoods of the points $\theta=\theta_0$ and $\theta=\pi -\theta_0$
are badly projected on the configuration space $(\theta,\phi)$. However, the Maslov
theory guarantees that it is possible to choose other local coordinates
such that the mapping from the Lagrangian manifold to the selected coordinates
is locally a diffeomorphism. In the present case it is easy to see that the mapping
in the plane $(p_\theta,\phi)$ is diffeomorphic (see Fig. 16 of Ref. \cite{Maslov1}). 
Once the local asymptotic solution in terms of $(p_\theta,\phi)$ has been computed,
it is possible to return to the configuration space $(\theta, \phi)$ by transforming
$p_\theta \rightarrow \theta$ through an inverse Fourier transform $F^{-1}_{k,p_\theta\rightarrow\theta}$.
It is indeed the asymptotic evaluation (for large values of $k$) of $F^{-1}_{k,p_\theta\rightarrow\theta}$,
obtained again by the stationary point method, that gives rise to an additional phase-shift 
in the solution. In fact, the term 
$\exp\{i(\pi/4)\,\Sg(\partial^2\Phi/\partial\beta^2)\}$
of formula (\ref{venti}) is modified as follows. Instead of $\Phi(\theta,\phi)$, it must
be considered the phase $\tilde{\Phi}(p_\theta,\phi)$ in the mixed representation coordinate-momentum; 
to have the phase described in the configuration space we move back from $p_\theta$ to
$\theta$ by means of the inverse Fourier transformation. Then, we have an exponential term
of the form $\exp\{ik[\tilde{\Phi}+\theta p_\theta]\}$ into an integral where $\theta$ is regarded
as a fixed parameter, whereas $p_\theta$ is the integration variable. 
Let $\Phi^{'}=\tilde{\Phi}+\theta p_\theta$.
The stationary point is then determined by the equality
\beq
\label{trentatre}
\frac{\partial\Phi^{'}}{\partial p_\theta}=\frac{\partial\tilde{\Phi}}{\partial p_\theta}
+\theta = 0,
\eeq
\noindent
and, therefore, it is given by that value of $p_\theta$ such that 
$\partial\tilde{\Phi} / \partial p_\theta = - \theta$.
Next, we get:
\beq
\label{trentacinque}
\Sg \frac{\partial^2\Phi^{'}}{\partial p_\theta^2} =
\Sg\frac{\partial^2\tilde{\Phi}}{\partial p_\theta^2} =
-\Sg\frac{\partial\theta}{\partial p_\theta}.
\eeq
\noindent
Therefore, the additional phase factor 
$\exp\{-i(\pi/4)\,\Sg (\partial\theta/\partial p_\theta)\}$ 
emerges.

The negative inertial index of a symmetric non--degenerate $(n\times n)$ matrix ${\cal A}$, called
{\it Inerdex}, is the number of negative
eigenvalues of the matrix \cite{Maslov1}. The following relationship holds \cite{Maslov1}:
$\Sg {\cal A} +2 \,\In {\cal A} = n$ (where $\Sg {\cal A}$ is for the signature of the quadratic
form associated with the matrix ${\cal A}$).
In our case we have
\beq
\label{trentasei}
-\Sg \frac{\partial\theta}{\partial p_\theta}=2 \In \frac{\partial\theta}{\partial p_\theta}-1,
\eeq
\noindent
then:
\beq
\label{trentasette}
\exp\left\{-i\,\frac{\pi}{4}\,\Sg \frac{\partial\theta}{\partial p_\theta}\right\} =
\exp\left\{i\,\frac{\pi}{2} \,\In \frac{\partial\theta}{\partial p_\theta} -i\frac{\pi}{4}\right\}.
\eeq
\noindent
We now focus our attention on the phase factor
$\exp\{i\delta\}=\exp\{i\,\frac{\pi}{2}\,\In(\frac{\partial\theta}{\partial p_\theta})\}$,
which is the relevant term in the analysis of the crossing through the critical points.
In close analogy with the Morse index which gives the dimension of the subspace
on which $E_{**}$ is negative definite,
and whose jumps count the increase of this dimension (see Section \ref{conjugate}), here it must be evaluated
the number of transitions from positive to negative values of 
$(\partial\theta/\partial p_\theta)$, or vice-versa. With this in mind,
we cover the curve: $p_\theta^2+(c_1^2/ \sin^2\theta)=1$, diffeomorphic
to a circle, with four charts: let ${\cal U}^1$ be the chart which lies
in the half-plane $p_\theta <0$, ${\cal U}^3$ be the chart which lies in the
half-plane $p_\theta > 0$, and ${\cal U}^2$ and ${\cal U}^4$ be the charts
which lie in the neighbourhoods of the points $(\theta_0,0)$ and
$(\pi -\theta_0,0)$, respectively. 
Next, the evaluation of $(\partial\theta/\partial p_\theta)$ gives
\beq
\label{trentotto}
\frac{\partial\theta}{\partial p_\theta}=\frac{c_1 p_\theta}
{(1-p_\theta^2)\sqrt{1-(p_\theta^2+c_1^2)}},~~~~(0 < c_1 < 1).
\eeq
\noindent
Let us consider a path $l$, counterclockwise oriented, which connects
two points $r'$ and $r''$ belonging to ${\cal U}^1$ and ${\cal U}^3$, respectively. 
The Maslov index $\Id (l)$, defined as the variation of $\In (\partial\theta/\partial p_\theta)$ 
along $l$, reads
\beq
\label{trentanove}
\Id (l)=\In \left (\frac{\partial\theta}{\partial p_\theta}\right )_{\!\! r''} -
\In \left (\frac{\partial\theta}{\partial p_\theta}\right )_{\!\! r'} = -1,
\eeq
\noindent
since $ \In \left (\partial\theta/\partial p_\theta\right )_{r''} = 0$,
because $p_\theta > 0$, while 
$\In \left (\partial\theta/\partial p_\theta\right )_{r'} = 1$, because $p_\theta < 0$.
Accordingly, we have a phase-factor: $\exp\{i\delta\}=\exp\{-i\pi/2\}$.
Conversely, for any other path $l'$ connecting two points lying in the same chart
${\cal U}^3$ (or ${\cal U}^4$) that does not intersect the cycle of the
singularities, we have $\Id (l')=0$.

\begin{remark}
\rm
Unfortunately there is a very unpleasant ambiguity concerning
the sign of $\Id (l)$ (see Ref. \cite{Delos}). Here we follow the Maslov prescription
\cite{Maslov2}: the transition through a critical (focal) point in the direction
of decreasing $(\partial\theta/\partial p_\theta)$ increases the index $\Id (l)$ 
by one; the transition in the direction of increasing
$(\partial\theta/\partial p_\theta)$ decreases the index $\Id (l)$ by one.
\end{remark}

Now, let us consider the patchwork across the critical points for
the derivation of the {\it connection-formulae}. As we have seen previously,
$u(k,\theta,\phi)$ can be approximated, for large values of $k$, by the 
expression: $A\exp\{ik\Phi_\pm\}$ (see formula (\ref{trentuno})). After
crossing a critical point, say $(\theta_0,0)$, an additional $-\pi/2$ phase-shift
arises, and, accordingly, the solution becomes $A\exp\{i[k\Phi_\pm - \pi/2]\}$.
By taking the limit $c_1\rightarrow 0$, we obtain the following 
connection-formulae:
\begin{enumerate}
\item[a)] For the rays traveling around the sphere in counterclockwise sense,
we have:
\end{enumerate}
\beq
\label{quaranta}
e^{ik\theta}~~ \stackrel{\AC}{\displaystyle \put(0,0){\vector(1,0){40}}}
~~e^{i\,[k\theta -\frac{\pi}{2}]},
\eeq
\noindent
\begin{enumerate}
\item[]
(we use the convention of taking positive the arcs counterclockwise oriented).
\item[b)] For the rays traveling around the sphere in clockwise sense, we have:
\end{enumerate}
\beq
\label{quarantuno}
e^{-ik\theta}~~ \stackrel{\AC}{\displaystyle \put(0,0){\vector(1,0){40}}}
~~e^{-i\,[k\theta -\frac{\pi}{2}]},
\eeq
\noindent
\begin{enumerate}
\item[]
(the arcs oriented in clockwise sense are taken negative).
\end{enumerate}

Finally, in view of the homotopic invariance of the Maslov index \cite{Maslov1},
the bijective correspondence between the winding number, associated with the
fundamental group $\pi_1 (X;p_0)$, and the {\it crossing-number} (discussed
in Section \ref{morse}) can now be extended to the Maslov phase-shift
as follows:
\begin{eqnarray}
\Z & ~~~\longrightarrow ~~~& ~~~~ \N \hspace{0.11cm}~~~~~\longrightarrow ~~~\mbox{Maslov phase-shift} \nonumber \\
m        & ~~~\longrightarrow ~~~& (2m-1) ~\longrightarrow ~~~~~~-\frac{\pi}{2} \, (2m-1) \hspace{1cm} (m>0) 
\label{quarantaduea} \\
m        & ~~~\longrightarrow ~~~& ~-2m~~~~\longrightarrow ~~~~~~~~~\,\frac{\pi}{2} \, (-2m) \hspace{1.4cm} (m \leq 0)
\label{quarantadueb}
\end{eqnarray}
In particular, notice that, for every complete tour around the sphere, 
both the counterclockwise and the clockwise
oriented rays acquire a factor $(-1)$, due to the product of two Maslov factors.

\begin{remark}
\rm
i) In the literature the term {\it creeping-waves}
usually indicates waves creeping along the boundary and continuously sheding energy
into the surrounding space. This is also the meaning of this term in the present paper.
However, only in the present subsection, this term has been used with a slight different
meaning: the waves are creeping around the obstacle, but they are supposed not to irradiate
around. In fact, in this subsection we have considered a manifold without boundary.
In the next subsection the problem will be reconsidered in its full generality
as an obstacle problem in a Riemannian manifold with boundary, and we shall evaluate
the damping factors associated with the rays that propagate
along the boundary, and that shed energy into the surrounding space.\\
ii) Guillemin and Sternberg, in their excellent book \cite{Guillemin} on Geometric Asymptotics,
give an analysis of Maslov's indexes, and illustrate the related application in a very general setting, with 
particular attention to the geometric quantization. They also calculate the $\pi/2$ phase--shift
at the crossing of the caustic by the use of the Morse theory.
\end{remark}

\subsection{Airy Approximation and Damping Factors}
\label{airy}
The diffraction problem concerns with the determination of a solution
$u_s(\bx,k)$ of the reduced wave equation (\ref{quindici})
in the exterior of a closed surface $S$ that, in our case, is a sphere of radius $R$
(embedded in $\R^3$). 
An incident field $u_i(\bx,k)$ that satisfies Eq. (\ref{quindici}) is prescribed,
and $u_s(\bx,k)$ is also required to satisfy the radiation condition
\beq
\label{quarantatre}
\lim_{|\bx|\rightarrow\infty} |\bx| \left (\frac{\partial}{\partial |\bx|} u_s -i\,k\,u_s\right ) =0,
\eeq
\noindent
(where $|\bx | = (\sum_{i=1}^3 x_i^2)^{1/2}$, $x_i$ being the cartesian coordinates of the
ambient space $\R^3$). Finally, the total field $u(\bx,k)$ must also satisfy a boundary 
condition on the obstacle (see below formula (\ref{cinquantadue})).

In order to solve this problem we intend to apply once again the method of the
stationary phase to an integral of the form (\ref{diciassette}). But, in this case,
the surface $S$ (boundary of a Riemannian manifold $\MMS$) is the envelope of the
diffracted rays: it is a caustic. We shall prove in the next subsection that the 
approximation given by eqs. (\ref{ventuno}) and (\ref{ventidue}) fails locally on this
domain, since the amplitude becomes infinite.
Each point $P$ outside the caustic lies on the intersection
of two diffracted rays which are tangent to the border of the diffracting ball, and 
crossing orthogonally two surfaces of constant phase $\Phi^\pm$ 
(for simplicity, we are now considering rays that are not bending around the obstacle). 
When the point $P$ is pushed on the boundary, the curves of constant phase meet forming a cusp, and,
accordingly, the stationary points $\beta_1$ and $\beta_2$ of the two phase functions $\Phi^\pm$
coalesce. In such a situation the standard method,
used in Section \ref{creeping} fails, and a different strategy must be looked for.
An appropriate procedure is the one suggested
by Chester, Friedmann and Ursell \cite{Chester}, that consists 
in bringing the phase function $\Phi$ into a more convenient form
by a suitable change of the integration variable $\xi \leftrightarrow \beta$, implicitly defined by
\beq
\label{quarantaquattro}
\Phi(\bx,\beta)= \Theta_0(\bx)+ F(\bx,\xi).
\eeq
\noindent
After this change, integral (\ref{diciassette}) reads
\beq
\label{quarantacinque}
u(\bx,k)=e^{ik\Theta_0(\bx)} \int g(\bx,\xi)~e^{ikF(\bx,\xi)}~d\xi.
\eeq
\noindent
This expression is indeed similar to the integral (\ref{diciassette}),
with the phase-function $\Phi$ replaced by $F$, and with an additional oscillatory factor
$\exp\{ik\Theta_0\}$ in front.

\noindent
Now, we have two stationary points $\beta_1(\bx)$ and $\beta_2(\bx)$ that coalesce, 
and our goal is to choose a transformation such that
to these points there correspond the points where $\partial F/\partial\xi$ 
vanishes. This result can be achieved by setting
\beq
\label{quarantasei}
F(\bx,\xi)=-\frac{1}{3}\xi^3+ \rho_0(\bx)\, \xi.
\eeq
\noindent
In fact, $\partial F/\partial\xi=-\xi^2+\rho_0(\bx)=0$ gives $\xi=\pm\sqrt\rho_0$. Then, from 
equalities (\ref{quarantaquattro}) and (\ref{quarantasei})
we obtain the following relationships:
\begin{eqnarray}
\sds \Phi(\beta_1,\bx) & = & \frac{2}{3}\,\rho_0^{3/2}(\bx) + \Theta_0(\bx), \label{quarantasettea} \\
\sds \Phi(\beta_2,\bx) & = & -\frac{2}{3}\,\rho_0^{3/2}(\bx) + \Theta_0(\bx), \label{quarantasetteb}
\end{eqnarray}
\noindent
that yield
\begin{eqnarray}
\sds \Theta_0(\bx) & = & \frac{1}{2} \left\{\Phi(\beta_1,\bx)+\Phi(\beta_2,\bx)\right\}, \label{quarantottoa} \\
\sds \frac{2}{3}\,\rho_0^{3/2}(\bx) & = & \frac{1}{2} \left\{\Phi(\beta_1,\bx)-\Phi(\beta_2,\bx)\right\}. 
\label{quarantottob}
\end{eqnarray}
In the case $\beta_1 = \beta_2$, we have $\rho_0(\bx) = 0$ and 
$\Theta_0(\bx) = \Phi(\beta_1,\bx)=\Phi(\beta_2,\bx)$.
If equalities (\ref{quarantottoa}, \ref{quarantottob}) are satisfied, 
the transformation $\xi \leftrightarrow \beta$
is uniformly regular and $1-1$ near $\xi=0$ (see Ref. \cite{Chester}). 
From eqs. (\ref{quarantacinque}) and (\ref{quarantasei}) it follows that the most significant terms
in the expression of $u(\bx,k)$, for large $k$, can be written in terms of
the Airy function $\Ai(\cdot)$ and of its derivative $\Ai'(\cdot)$.
Such a representation of $u(\bx,k)$ led Ludwig \cite{Ludwig} to propose the following
{\it Ansatz}:
\beq
\label{cinquantuno}
u(\bx,k)= e^{ik\Theta_0(\bx)}\left\{g_0(\bx) \Ai(-k^{2/3}\rho_0) +
ik^{-1/3} h_0(\bx) \Ai' (-k^{2/3}\rho_0)\right\},
\eeq
where $g_0(\bx)$ and $h_0(\bx)$ are respectively the first terms of the two formal 
asymptotic series:\\ $\sum_{m=0}^\infty g_m(\bx)k^{-m}$, and 
$\sum_{m=0}^\infty h_m(\bx) k^{-m}$.

Unfortunately, representation
(\ref{cinquantuno}) cannot be made to satisfy the boundary condition at the surface
of the obstacle. In fact, on any point of the surface $S$ the two stationary points 
coalesce, and the phase of the diffracted ray can be assumed to coincide 
with the phase of the incident ray, that is $\Phi(\beta_1,\bx)=
\Phi(\beta_2,\bx)=\Phi_i(\bx)=$ phase of the incident wave. Consequently, $\rho_0(\bx)$ is
identically zero on $S$ (see Eq. (\ref{quarantottob})). It is therefore necessary to
introduce an appropriate modification of the asymptotic representation (\ref{cinquantuno})
in order to satisfy a boundary condition which, in its general form, reads \cite{Lewis}:
\beq
\label{cinquantadue}
\frac{\partial u}{\partial N} +i\,k^{2/3}\,\zeta \,u = 0 ~~~~~(\mbox{on}\,S),
\eeq
\noindent
where $\zeta$ is a smooth impedence function defined on $S$, and
$\partial u/\partial N$ is the normal derivative. For $\zeta=0$, and
$\zeta=\infty$ the relationship (\ref{cinquantadue}) reduces respectively
to the Neumann's $\partial u/\partial N=0$, and to the Dirichlet's $u=0$
boundary condition on $S$. Following Lewis, Bleinstein and Ludwig \cite{Lewis},
we introduce the new {\it Ansatz}, which is a slight modification of the
asymptotic representation (\ref{cinquantuno}):
\beq
\label{cinquantatre}
u(\bx,k)= e^{ik\Theta(\bx)}\left\{g(\bx) A(-k^{2/3}\rho) +
ik^{-1/3} h(\bx) A'(-k^{2/3}\rho)\right\},
\eeq
\noindent
where:
\begin{eqnarray}
\sds g(\bx) & \sim & \sum_{m=0}^\infty\frac{g_m(\bx)}{k^{m/3}},~~~~~ h(\bx)\sim\sum_{m=0}^\infty\frac{h_m(\bx)}{k^{m/3}},
\label{cinquantaquattroa} \\
\sds\rho(\bx) & = & \rho_0(\bx)+k^{-2/3}\rho_1(\bx),~~~~~(\rho_0(\bx)=0 ~\mbox{on}~ S), \label{cinquantaquattrob} \\
\sds \Theta(\bx) & = & \Theta_0(\bx) + k^{-2/3} \Theta_1(\bx), \label{cinquantaquattroc} \\
\sds A(t) & = & \Ai (te^{\,i\,2\pi/3}). \label{cinquantaquattrod}
\end{eqnarray}
Now, by inserting formula (\ref{cinquantatre}) into the reduced wave equation (\ref{quindici}),
and collecting the coefficients of $k^{m/3}A \exp\{ik\Theta\}$ and
$k^{m/3}A' \exp\{ik\Theta\}$, we obtain the following set of equations:
\begin{eqnarray}
&&(\nabla\Theta_0)^2+\rho_0 (\nabla\rho_0)^2=1, \label{cinquantacinquea} \\
&&\nabla\Theta_0\cdot\nabla\rho_0 = 0, \label{cinquantacinqueb} \\
\label{cinquantacinquec}
&&2\nabla\Theta_0\cdot\nabla\Theta_1+\rho_1(\nabla\rho_0)^2+2\rho_0\nabla\rho_0\cdot\nabla\rho_1=0, \\
\label{cinquantacinqued}
&&\nabla\Theta_0\cdot\nabla\rho_1+\nabla\Theta_1\cdot\nabla\rho_0 = 0.
\end{eqnarray}
\noindent
Then the {\it Ansatz} (\ref{cinquantatre}) can be inserted into the boundary
condition (\ref{cinquantadue}). Here we limit ourselves to report the result which is
relevant in our case considering, for simplicity, the Dirichlet boundary condition,
i.e. putting $\zeta=\infty$ in formula (\ref{cinquantadue}).
The reader interested to the details of the calculations is referred to Ref. \cite{Lewis}.
It can be easily seen from Eqs. (\ref{cinquantacinquec}) and (\ref{cinquantacinqued}) that $\Theta_1$ acquires an imaginary term, 
that depends on the curvature of the obstacle.
If the obstacle is a sphere of radius $R$, we have:
\beq
\label{sessantasette}
\Theta_1^{(i)} = \Cs - \frac{1}{2} \, q_i  \, e^{\,i\pi/3}\left (\frac{2}{R}\right )^{2/3}
\int_0^l\: d\tau,
\eeq
$l$ being the length of the arc on the surface of the sphere described by the diffracted rays,
and $q_i$ denoting the $i$-th zero of the Airy function. 
For each $q_i$ one gets a solution (\ref{cinquantatre}) of the reduced wave equation with
boundary condition (\ref{cinquantadue}), and, correspondingly, an infinite set of
{\it damping--factors} $\alpha_i$ given by
\beq
\label{sessantanove}
\alpha_i=e^{-\beta_i l},~~~\beta_i= \Cs |q_i| \, k^{1/3} \, R^{-2/3},
\eeq
\noindent
which depend on the curvature of the obstacle.

\begin{remark}
\rm
In connection with the results of this subsection we want to mention
the deep and extensive works of Berry \& Howls \cite{Berry4,Berry5}, and Connor and 
collaborators \cite{Connor3} on the asymptotic evaluation of oscillating integrals 
when two or more saddle points coalesce. 
Moreover, it is worth mentioning the paper by Berry \cite{Berry3}, where the first
general statement of uniform approximations using catastrophe theory has been done
(for the application to special cases see also Ref. \cite{Connor2}).
\end{remark}

\begin{figure}[htb]
\vspace{-5in}
\centering
\includegraphics[width=9in]{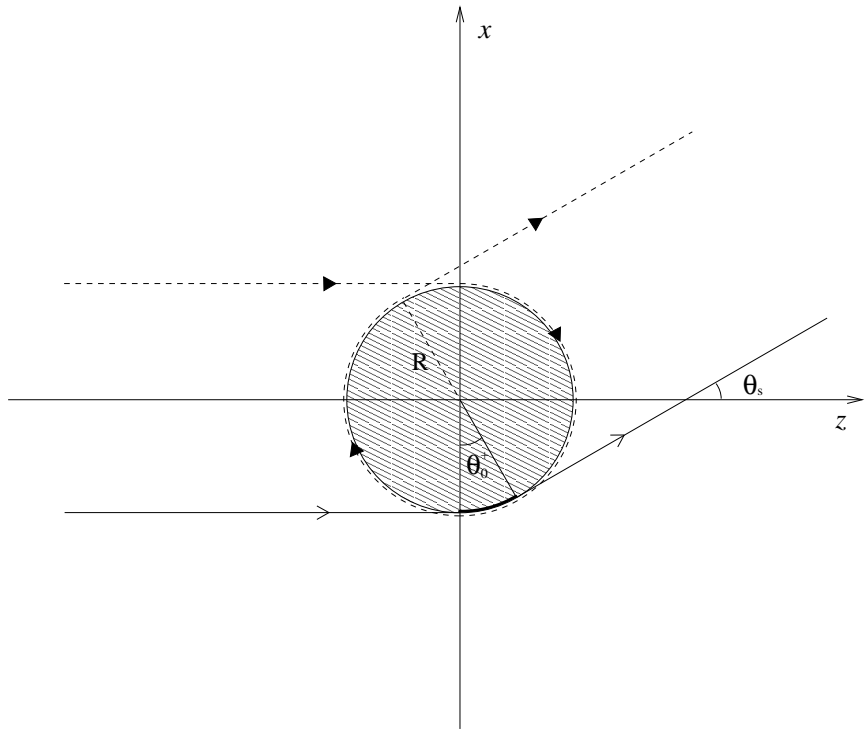}
\caption{
Diffractive scattering: geometry of the contribution of two grazing rays to the scattered amplitude.
The ray below (solid line with open arrows) travels $\theta_0^+$ radians in the counterclockwise direction along
the boundary of the obstacle, and leaves
the obstacle in the direction of the scattering angle $\theta_s$ (see formula (\ref{ottantuno})).
The ray above (solid and dashed lines with filled arrows) travels in the
clockwise direction along the surface of the obstacle and crosses the axial caustic twice before
emerging in the direction of the scattering angle $\theta_s$ (see formula (\ref{ottantadue})).
}
\label{figura_3}
\end{figure}

\subsection{Surface Waves in Diffractive Scattering}
\label{scattering}
In the scattering theory the far field diffracted by the obstacle must be related to the incoming
field whose source is located at great distance from the obstacle. Then, we have to 
consider the limit obtained when the source point $p_0$ is pushed to $-\infty$ and the 
observer to $+\infty$.
Let us introduce an orthogonal system of axes $(x,y,z)$ in $\R^3$ whose origin coincides with the center
of the sphere, and such that the $z$-axis, chosen parallel to the incoming beam of rays, is 
positively oriented in the direction of the outgoing rays (see Fig. \ref{figura_3}). 
Let us introduce, in addition, the coordinates
on the sphere: ${\bf r}_0$ is the radial vector, $\phi_0$ is the azimuthal angle, $\theta_0$
is the angle measured along the meridian circle from the point of incidence of the ray on the sphere.
Finally, let $\tau$ be a parameter running along the ray; in 
particular we have (on the sphere): $\tau=R\theta_0$ ($R$ being the radius of the sphere). 
Now, consider the rays that leave the surface of the sphere after diffraction. In view of the fact that the
interior of the space, outside the obstacle, can be regarded as a Riemannian euclidean space
without boundary, eqs. (\ref{ventiquattroa}, \ref{ventiquattrob}) hold true, and, in this case, read 
\begin{eqnarray}
\sds \frac{d{\bf r}}{d\tau} & = & \PP, \label{settantunoa} \\
\sds \frac{d\PP}{d\tau} & = & 0, \label{settantunob}
\end{eqnarray}
\noindent
where ${\bf r}=(x,y,z)$, and $\PP=\nabla\Phi$ ($\Phi=$ phase function). 
From eqs. (\ref{settantunoa}, \ref{settantunob}) we get:
\beq
\label{settantadue}
{\bf r}(\theta_0,\phi_0,\tau)={\bf r}_0(\phi_0,\theta_0)+\tau\PP_0 (\phi_0,\theta_0),
\eeq
\noindent
where $\PP_0$ is the unit vector tangent to the obstacle where the ray leaves the sphere. 

\noindent
Let us now focus our attention on the ray hitting the sphere at the point of coordinates $(-R,0,0)$, 
and then traveling in counterclockwise sense.
The components of the radial vector are (see Fig. \ref{figura_3}): 
${\bf r}_0=(-R\cos\theta_0\cos\phi_0, R \cos\theta_0 \sin\phi_0,R\sin\theta_0)$,
and 
$\PP_0=(\sin\theta_0\,\cos\phi_0, -\sin\theta_0\sin\phi_0,\cos\theta_0)$.
Substituting these expressions in Eq. (\ref{settantadue}) we have:
\beq
\label{settantatre}
{\bf r}(\theta_0,\phi_0,\tau) = (-R\cos\theta_0\cos\phi_0+\tau\sin\theta_0\cos\phi_0,
R\cos\theta_0\sin\phi_0-\tau\sin\theta_0\sin\phi_0, R\sin\theta_0+\tau\cos\theta_0).
\eeq
\noindent
Now, the domain where the Jacobian 
$J={\sds\frac{\partial(x,y,z)}{\partial(\theta_0,\phi_0,\tau)}}=\tau(R\cos\theta_0-\tau\sin\theta_0)$
vanishes is composed by:
\begin{enumerate}
\item[i)] the surface of the sphere, where $\tau=0$;
\item[ii)] the semi-axis represented by $\tau=\bar{\tau}=R\,\cot\,\theta_0$.
\end{enumerate}
We can rewrite the transport equation (\ref{ventidue}) in the following form \cite{Maslov1}:
\beq
\label{settantacinque}
\frac{1}{J}\frac{d}{d\tau} (JA^2) = 0,
\eeq
\noindent
whose solution $A=C / \sqrt J$ indicates once again that the amplitude becomes infinite for $J=0$,
i.e. on the caustic.
In order to treat the scattering problem we must perform the limit for $r\rightarrow +\infty$ ($r=|{\bf r}|$).
Since $\tau=\sqrt{r^2-R^2}$, then as $r\rightarrow\infty$, $\tau$ tends to $r$, and
from the expression of the Jacobian, we obtain: 
$\sqrt J \staccrel{\displaystyle\longrightarrow}{r\rightarrow +\infty} i\,r\,\sqrt{\sin\theta_0}$.

Now, let us revert to formula (\ref{cinquantatre}) in order to evaluate the contribution
due to the diffracted ray. Recalling the asymptotic behaviour of
$\Ai$ and $\Ai'$, and by noting that in the present geometry $\Theta_0 = R\theta_0+\tau$, and that
the length of the arc described by the diffracted ray on the sphere is $l=R\theta_0$, we obtain,
for $k\rightarrow +\infty$ and for large values of $r$ (see formula (\ref{sessantanove}) 
and Ref. \cite{Lewis}):
\beq
\label{settantotto}
u_i(k,\theta_0,r)
\staccrel{\longrightarrow}{\staccrel{\scriptstyle k\rightarrow\infty}{\scriptstyle r\rightarrow\infty}}
C_i^{(d)}(k)\, \frac{e^{ikr}}{r}\, \frac{e^{-\beta_i R\,\theta_0} e^{ikR\,\theta_0}}{i\sqrt{\sin\theta_0}},
\eeq
$(0<\theta_0<\pi)$, where $\beta_i$ is the exponent of the $i$-th damping factor
$\alpha_i$ obtained in Section \ref{airy}. For what concerns the diffraction coefficients
$C_i^{(d)}(k)$, they have been derived by Lewis, Bleinstein and Ludwig \cite{Lewis}
by assuming that the amplitude of the diffracted wave is proportional to the amplitude
of the incident wave. In the present geometry these coefficients read
\beq
\label{settantanove}
C_i^{(d)}(k)=\Cs \frac{(kR)^{1/3}}{[\Ai' (q_i)]^2},
\eeq
($q_i$ being the zeros of the Airy function), and coincide with the expression derived by 
Levy-Keller (see Table I of Ref. \cite{Levy}).
Since the roots of the Airy function are infinite, we have, correspondingly, a countably
infinite set of modes $u_i$. For the moment we focus our attention on the damping factor 
$\alpha_0$, whose exponent $\beta_0$ is the smallest one, and, accordingly, on the mode $u_0$ that,
for large values of $k$ and $r$, reads
\beq
\label{ottanta}
u_0(k,\theta_0,r)
\staccrel{\longrightarrow}{\staccrel{\scriptstyle k\rightarrow\infty}{\scriptstyle r\rightarrow\infty}}
-i\, C_0^{(d)}(k)\, \frac{e^{ikr}}{r}\frac{e^{i\lambda_0\theta_0}}{\sqrt{\sin\theta_0}},
\eeq
$(0<\theta_0<\pi,\,\lambda_0=R\,(k+i\beta_0))$. 

In order to evaluate the contribution to the 
scattering amplitude, the scattering angle $\theta_s$ must be related to
the surface angle $\theta_0$. To distinguish between the contribution of the
counterclockwise rays and that of the clockwise ones, we add the superscript '$+$' to all what
refers to the counterclockwise rays, and the superscript '$-$' to all the symbols
referring to the clockwise oriented rays\footnote{In the present geometry the simple convention
adopted in Section \ref{creeping} is ambiguous and cannot be used. In the
present case the rays turning in clockwise sense hit the sphere at a point which
is antipodal with respect to the point where the rays, turning in counterclockwise
sense, hit the obstacle.}.
With this convention, and observing that
$\theta_s=\theta_0^+$ (see Fig. \ref{figura_3}), the contribution to the scattering amplitude
$f^+_{(0,0)}$ of the counterclockwise grazing ray that has not completed one
tour around the obstacle\footnote{In the notation $f^+_{(0,0)}$, the
first zero of the subscript indicates that the grazing ray has not completed one tour, whereas
the second one indicates that we are taking the smallest exponent $\beta_0$.}
can be written as
\beq
\label{ottantuno}
f^+_{(0,0)}(k,\theta_s)=C_0^{(d)}(k)\,\frac{e^{i\lambda_0\theta_s} e^{-i\pi/2}}
{\sqrt{\sin\theta_s}},~~~~~(0<\theta_s<\pi).
\eeq
\noindent
Notice that in formula (\ref{ottantuno}) the factor $\exp\{-i\pi/2\}$ corresponds to the Maslov 
phase-shift due to the fact that the ray crosses the axial caustic once (see Fig. \ref{figura_3}).

\noindent
Analogously, the contribution to the scattering amplitude $f^-_{(0,0)}$ of a diffracted
ray which travels around the sphere, in clockwise sense, without completing one
tour around the obstacle (see Fig. \ref{figura_3}) can be evaluated. To this purpose, it 
is convenient to consider the asymptotic limit of $|J|^{1/2}$,
which is given by $r\,|\sin\theta_0^-|^{1/2}$. Since the scattering angle
$\theta_s$ is related to the surface angle $\theta_0^-$ as follows: $\theta_0^- = 2\pi -\theta_s$,
and by noting that this (clockwise oriented) grazing ray crosses the
axial caustic (i.e. the $z$-axis) two times (see Fig. \ref{figura_3}), we have:
\beq
\label{ottantadue}
f^-_{(0,0)}(k,\theta_s)=(-1)\,C_0^{(d)}(k)\,\frac{e^{i\lambda_0(2\pi-\theta_s)}}{\sqrt{\sin\theta_s}},
~~~~~(0<\theta_s<\pi),
\eeq
\noindent
where the factor $(-1)$ is precisely given by the product of two Maslov 
factors\footnote{It is easy to prove, through arguments based on symmetry, that the
negative $z$-axis is a caustic for the rays propagating in the backward emisphere.} 
(see Section \ref{creeping}). \\
Adding $f^+_{(0,0)}$ to $f^-_{(0,0)}$, we obtain the scattering amplitude $f_{(0,0)}(k,\theta_s)$ 
due to the rays not completing one tour around the obstacle:
\beq
\label{ottantatre}
~~f_{(0,0)}(k,\theta_s)=f^+_{(0,0)}+f^-_{(0,0)}=
-i \, C_0^{(d)}(k) \,\frac{e^{i\lambda_0\theta_s}-i e^{i\lambda_0(2\pi-\theta_s)}}{\sqrt{\sin\theta_s}},
~~~~~(0<\theta_s<\pi).
\eeq
\noindent
Now, we are ready to take into account the contribution of all those rays which are orbiting 
around the sphere several times. Let us consider the rays describing $n$ ($n\in\N$)
tours around the obstacle. Since the surface angles $\theta_{0,n}^\pm$ are related
to the scattering angle $\theta_s$ as
$\theta_{0,n}^+ = \theta_s+2\pi n$, $\theta_{0,n}^- = 2\pi - \theta_s+2\pi n$ $(n=0,1,2,\ldots)$,
we have for $0<\theta_s<\pi$:
\beq
\label{ottantacinque}
f_0(k,\theta_s)=-i\, C_0^{(d)}(k)\sum_{n=0}^\infty (-1)^n\, e^{i2\pi n\lambda_0}\, 
\frac{e^{i\lambda_0\theta_s}-i e^{i\lambda_0(2\pi-\theta_s)}}{\sqrt{\sin\theta_s}}.
\eeq
\noindent
The factor $(-1)$, at each $n$, is due to the product of two Maslov phase factors, 
corresponding to the fact that both the counterclockwise and the clockwise
rays cross the $z$-axis (i.e. the axial caustic) twice for each tour (see Section \ref{creeping}).

\noindent
Now, we exploit the following expansion:
\beq
\label{ottantasei}
\frac{1}{2\cos\pi\lambda_0}=e^{i\pi\lambda_0}\sum_{n=0}^\infty (-1)^n e^{i2\pi n\lambda_0},
~~~~~(\mbox{Im}\,\lambda_0 > 0).
\eeq
\noindent
By the use of formula (\ref{ottantasei}), the amplitude (\ref{ottantacinque}) can be rewritten as
\beq
\label{ottantasette}
~~f_0(k,\theta_s)=-C_0^{(d)}(k)\,e^{i\pi/4}\,
\frac{e^{-i\,\{\lambda_0(\pi-\theta_s)-\pi/4\}}+e^{\,i\,\{\lambda_0(\pi-\theta_s)-\pi/4\}}}
{(2\cos\pi\lambda_0)\: \sqrt{\sin\theta_s}},~~~~~(0<\theta_s<\pi).
\eeq
\noindent
The r.h.s. of formula (\ref{ottantasette}) contains the asymptotic behaviour,
for $|\lambda_0|\rightarrow\infty$ and $|\lambda_0|(\pi-\theta_s)>>1$, of the Legendre function 
$P_{\lambda_0-\frac{1}{2}}(-\cos\theta_s)$ times $\sqrt{2\pi\lambda_0}$ \cite{Erdelyi}.
Then, writing
$P_{\lambda_0-\frac{1}{2}}(-\cos\theta_s)$ in place of its asymptotic behaviour, we have for
$|\lambda_0|\rightarrow\infty$ $(0<\theta_s \leq \pi)$:
\beq
\label{ottantotto}
f_0(k,\theta_s)\simeq -C_0^{(d)}(k)e^{i\pi/4}\,
\frac{\sqrt{2\pi\lambda_0}\, P_{\lambda_0-\frac{1}{2}}(-\cos\theta_s)}{2\cos\pi\lambda_0},
\eeq
that, by putting $\mu_0 = \lambda_0-1/2$, becomes
\beq
\label{ottantanove}
f_0(k,\theta_s)\simeq C_0^{(d)}(k)\,e^{i\pi/4}\,\frac{\sqrt\pi}{2}\,\sqrt{(2\mu_0+1)}\,
\frac{P_{\mu_0}(-\cos\theta_s)}{\sin\pi\mu_0}.
\eeq
\noindent
Now, if we consider the countable infinity of damping factors $\alpha_i$
(see Section \ref{airy}), we obtain an infinite set of creeping waves, whose angular
distribution is described by the Legendre functions $P_{\mu_i}(-\cos\theta_s)$, where
$\mu_i=\lambda_i-1/2,~\lambda_i=R(k+i\beta_i)$.

Let us recall that the simplest approach to diffraction scattering by a sphere is the
expansion of the scattering amplitude in terms of spherical functions. But, when
the radius of the sphere is large compared to the wavelength, these series converge so
slowly that they become practically useless (see Refs. \cite{Sommerfeld,Watson}). A typical
example of this difficulty is given by the diffraction of the radio waves around the earth.
In order to remedy this drawback, Watson \cite{Watson} proposed a resummation
of the series that makes use of the analytic continuation from integer values $l$ of the
angular momentum to complex $\lambda$-values (or $\mu$-values in our case). In this method
the sum over integral $l$ is substituted by a sum over an infinite set of poles
corresponding to the infinite set of creeping waves, whose angular distribution is described by the 
Legendre functions $P_{\mu_i}(-\cos\theta_s)$.
Let us note that at $\theta_s = 0$, $P_{\mu_i}(-\cos\theta_s)$ presents a
logarithmic singularity \cite{Sommerfeld}, and, consequently, approximation (\ref{ottantanove}) fails.
Furthermore, at small angles the surface waves describe a very small arc of
circumference, and the damping factors $\alpha_i$ are close to 1; therefore, we are
obliged to take into account the contribution of the whole set of creeping waves.
On the other hand, at $\theta_s = \pi$, $P_{\mu_i}(-\cos\theta_s) = 1$ and, furthermore,
at large angles, the main contribution comes from that creeping wave whose damping 
factor $\alpha_0=\exp(-\beta_0 l)$ has the smallest exponent $\beta_0$.
Therefore, in the backward angular region the surface wave contribution is dominant,
and the scattering amplitude can be approximately represented as follows:
\beq
\label{novanta}
f(k,\theta_s)\simeq f_0(k,\theta_s)={\cal G}_0(k)\, P_{\mu_0}(-\cos\theta_s),
~~~~~(\theta_s > 0),
\eeq
where ${\cal G}_0(k)=C_0^{(d)}(k) \, e^{i\frac{\pi}{4}}\, (\frac{\sqrt\pi}{2}) \, (2\mu_0+1)^{1/2} / \sin\pi\mu_0$. \\

\section{Conclusions}
\label{conclusions}
Let us conclude with a brief remark on the difference between complex and 
diffracted rays. Complex rays are well known in optics in total reflection, where they
describe the exponentially damped penetration into the rarer medium associated with
surface waves traveling along the boundary. Then, it is very tempting
to describe diffraction in terms of complex rays.
Conversely, in the theory of diffraction which we present in this paper, {\it we do not
make any use of complex rays, but rather we introduce the diffracted rays, that are real}.
The first difficulty which emerges in this approach is the proof of the existence
of these diffracted rays. This problem is solved by showing the non-uniqueness of the
Cauchy problem for the geodesics in a Riemannian manifold with boundary. 
Then, since the border of the obstacle is the envelope of the diffracted rays, and the standard
method of the stationary phase cannot be used,  we are forced to
apply a modified stationary phase method due to Chester, Friedman and Ursell \cite{Chester}.
By using this method we derive an infinite set of damping factors associated with the waves
creeping around the obstacle. These damping factors have again a geometrical nature, since they
depend on the curvature of the obstacle. In conclusion, it worth remarking that if we pass from optics to
mechanics, and we consider particle trajectories instead of light rays, the splitting
of geodesics at the boundary introduces a probabilistic aspect reflecting the fact that, at any point
of the boundary, the particle can continue to orbit around the obstacle or can
leave the obstacle itself. Therefore, using the damping factors is a way to connect probability,
proper of semi--classical mechanics, to geometry: i.e. the curvature of the obstacle.

\subsection*{Acknowledgments}
It is a pleasure to thank our friend Prof. M. Grandis for several helpful discussions.


\newpage

\begin{thebibliography}{32}

\baselineskip=12pt

\bibitem{Keller1}
J. B. Keller,
Proc. Symposia Appl. Math. {\bf 8} (1958) 27.

\bibitem{Keller2}
J. B. Keller,
J. Opt. Soc. Am. {\bf 52} (1962) 116.

\bibitem{Levy}
B. Levy and J. B. Keller,
Commun. Pure Appl. Math. {\bf 12} (1959) 159.

\bibitem{Bouche}
D. Bouche, F. Molinet and R. Mittra,
{\em Asymptotic Methods in Electromagnetics},
Springer--Verlag, Berlin, 1997.

\bibitem{Alexander2}
S. B. Alexander, I. D. Berg and R. L. Bishop,
Lecture Notes in Math. {\bf 1209} (1986) 1.

\bibitem{Alexander1}
S. B. Alexander, I. D. Berg and R. L. Bishop,
Illinois J. Math. {\bf 31} (1987) 167.

\bibitem{Berry1}
M. V. Berry and C. Upstill,
Prog. Optics {\bf 18} (1980) 257.

\bibitem{Milnor1}
J. Milnor,
{\em Morse Theory},
Princeton University Press, Princeton, NJ, 1963.

\bibitem{Maslov1}
V. P. Maslov and M. V. Fedoryuk,
{\em Semi-Classical Approximation in Quantum Mechanics},
D. Reidel Publishing, Dordrecht, 1981.

\bibitem{Mishchenko1}
A. S. Mishchenko, V. E. Shatalov and B. Y. Sternin,
{\em Lagrangian Manifolds and the Maslov Operator},
Springer-Verlag, Berlin, 1980.

\bibitem{Chester}
C. Chester, B. Friedman and F. Ursell,
Proc. Cambridge Phil. Soc. {\bf 54} (1957) 599.

\bibitem{Kravtsov1}
Y. A. Kravtsov,
Radiofizika {\bf 7} (1964) 664.

\bibitem{Ludwig}
D. Ludwig,
Commun. Pure Appl. Math. {\bf 19} (1966) 215.

\bibitem{Neubauer}
W. G. Neubauer,
``Observation of Acoustic Radiation from Plane and Curved Surfaces'',
pp. 61--126 in Physical Acoustics {\bf 10}, W. P. Mason and R. N. Thurston, eds.,
Academic Press, New York, 1973.

\bibitem{Fioravanti}
R. Fioravanti and G. A. Viano,
Phys. Rev. C {\bf 55} (1997) 2593.

\bibitem{Besse}
A. Besse,
{\em Manifolds all of whose Geodesics are Closed},
Springer--Verlag, Berlin, 1978.

\bibitem{Sommerfeld}
A. Sommerfeld,
{\em Partial Differential Equations in Physics},
Academic Press, New York, 1964.

\bibitem{Watson}
G. N. Watson,
Proc. Roy. Soc. London {\bf 95} (1918) 83.

\bibitem{Nussenzveig2}
H. M. Nussenzveig,
{\em Diffraction Effects in Semiclassical Scattering},
Cambridge University Press, Cambridge, 1992,
(see also the papers quoted therein).

\bibitem{Goldstein}
H. Goldstein,
{\em Classical Mechanics},
Addison-Wesley, Reading, 1959.

\bibitem{Plaut}
C. Plaut,
Compositio Mathematica {\bf 81} (1992) 337.

\bibitem{Kobayashi}
S. Kobayashi,
``On Conjugate and Cut Loci'', 
pp. 96--122 in Studies in Global Geometry and Analysis, S. S. Chern, ed.,
MAA Studies in Math. {\bf 4}, 1967.

\bibitem{Alexander3}
S. B. Alexander,
Springer Lecture Notes {\bf 838} (1981) 12.

\bibitem{Delos}
J. B. Delos,
Adv. Chem. Phys. {\bf 65} (1986) 161.

\bibitem{Maslov2}
V. P. Maslov,
{\em Operational Methods},
Mir Publishers, Moscow, 1973.

\bibitem{Guillemin}
V. Guillemin and S. Sternberg,
{\em Geometric Asymptotics}, in Mathematical Surveys {\bf 14}, 
American Mathematical Society, Providence, 1977.

\bibitem{Lewis}
R. M. Lewis, N. Bleinstein and D. Ludwig,
Commun. Pure Appl. Math. {\bf 20} (1967) 295.

\bibitem{Berry4}
M. V. Berry and C. J. Howls,
Proc. R. Soc. Lond. A {\bf 443} (1993) 107.

\bibitem{Berry5}
M. V. Berry and C. J. Howls,
Proc. R. Soc. Lond. A {\bf 444} (1994) 201.

\bibitem{Connor3}
J. N. L. Connor,  P. R. Curtis and R. A. W. Young,
``Uniform Asymptotics of Oscillating Integrals: Applications in Chemical Physics'',
pp. 24--38 in Wave Asymptotics, P. A. Martin and G. R. Wickham, eds.,
Cambridge University Press, Cambridge, 1992,
(see also the papers quoted therein).

\bibitem{Berry3}
M. V. Berry,
Adv. Phys. {\bf 25} (1976) 1.

\bibitem{Connor2}
J. N. L. Connor,
Molecular Phys. {\bf 31} (1976) 33.

\bibitem{Erdelyi}
A. Erdelyi, W. Magnus, F. Oberhettinger and F. G. Tricomi,
{\em Higher Trascendental Functions}, Vol. 1,
McGraw-Hill, New York, 1953.

\end{thebibliography}
\end{document}